\documentclass[preprint,prb,showpacs,floatfix]{revtex4}

\usepackage{graphicx}
\newcommand*{\cm}{cm$^{-1}$}
\newcommand{\comment}[1]{}
\newcommand*{\Co}{Na$_x$CoO$_2$\ }

\begin{document}

\title{The infrared conductivity of Na$_x$CoO$_2$: evidence of gapped states.}

\author{J. Hwang$^{1}$, J. Yang$^{1}$, T. Timusk$^{1,2}$ and  F.C. Chou$^{3}$}
\affiliation{$^{1}$Department of Physics and Astronomy, McMaster
University, Hamilton ON L8S 4M1, Canada \\ $^{2}$The Canadian
Institute of Advanced Research, Canada\\
$^3$Center for Materials Science and Engineering, MIT, Cambridge,
MA 02139, USA} \email{timusk@mcmaster.ca}

\begin{abstract}

We present infrared ab-plane conductivity data for the layered
cobaltate Na$_x$CoO$_2$ at three different doping levels ($x=0.25,
0.50$, and $0.75$). The Drude weight increases monotonically with
hole doping, $1-x$. At the lowest hole doping level $x$=0.75 the
system resembles the normal state of underdoped cuprate
superconductors with a scattering rate that varies linearly with
frequency and temperature and there is an onset of scattering by a
bosonic mode at 600 \cm. Two higher hole doped samples ($x=0.50$
and 0.25) show two different-size gaps (110 \cm and 200 \cm,
respectively) in the optical conductivities at low temperatures
and become insulators. The spectral weights lost in the gap region
of 0.50 and 0.25 samples are shifted to prominent peaks at 200 \cm
and 800 \cm, respectively. We propose that the two gapped states
of the two higher hole doped samples ($x$=0.50 and 0.25) are
pinned charge ordered states.

\end{abstract}


\maketitle

\comment{Introduction}

The recently discovered cobalt oxide superconductor
Na$_x$CoO$_2\cdot y$H$_2$O, with $1/ 4 < x < 1 / 3$ and
y=1.4\cite{takada03,schaak03,chou03,jin03}, resembles in many ways
that other famous family of oxide superconductors, the cuprates.
The cobaltate structure is based on two-dimensional planes, weakly
coupled in the third direction, the c-axis. The Co ions are on a
triangular lattice and calculations based on the t-J model suggest
that the material may give rise to novel quantum states
\cite{baskaran03,wang04}. The \Co material with $x=0.75$ appears
to be metallic with a Curie-Weiss susceptibility
\cite{ray99,foo04}. With increased hole doping, 1-$x$, (smaller
$x$) the \Co material shows a different metallic state with a
temperature independent paramagnetic susceptibility\cite{foo04}.
Separating the two metallic regions is an insulating state at
$x=0.50$ \cite {foo04}. In this state a superstructure develops in
the Na plane that separates the CoO planes and it has been
suggested that the insulating state is induced by this
superstructure \cite{foo04}. Other charge ordered states that
compete with superconductivity have been proposed at higher hole
doping levels than 0.50\cite{baskaran03a,koshibae03}. Here we
present optical data that show evidence of {\it two} gapped
insulating states, one at $x=0.50$ and a new one at a higher hole
doping level close to $x=0.25$.

We use optical spectroscopy to investigate the low-lying states of
the normal state of \Co at three different doping levels
($x=0.75$, 0.50, and 0.25) at various temperatures. Previous
optical work has been confined to the \Co materials with low hole
doping levels, $x$ above 0.50 ($x=0.57$, 0.70, and 0.82)
\cite{wang03,lupi03,bernhard04,caimi04}. A metallic Drude-like
absorption is found with anomalous linear frequency dependence of
the scattering rate. The infrared reflectance resembles that of
the cuprates with the reflectance decreasing linearly with
frequency over a broad region extending to nearly 0.75 eV. As the
temperature is lowered below room temperature, a knee develops in
the reflectance at 600 \cm, characteristic of the interaction of
the charge carriers with a bosonic mode
\cite{carbotte99,hwang04a}. We complement this work by extending
it to higher hole doping levels, examining electro-chemically
doped samples with $x=0.25$ and 0.50. We also studied an as-grown
sample with $x=0.75$.

The preparation of the samples has been described previously
\cite{chou03}. The parent \Co single crystals are grown by the
floating zone method. Na atoms which reside between the tilted
octahedral cobalt oxide layers were deintercalated in a
electrochemical cell. The Na concentration was determined by
electron probe microanalysis (EPMA) and we estimate the
uncertainty in $x$ to be $\pm$ 0.08. The hole doping of the cobalt
oxide layers is thought generally to be proportional to $1-x$ but
it should be noted that there have been reports of other sources
carriers such as oxygen deficiency\cite{chou04,banobre04} or
interlayer oxonium ion~\cite{takada04,milne04,chen05} making the
Na concentration a less reliable indicator of the in-plane doping
level. As we will show below a more direct measure of doping is
the Drude spectral weight which we will use as our main guide in
determining the planar hole concentration.

The optical experiments were performed on freshly cleaved ab-plane
surface of $2 \times 2$ mm$^2$ crystals in a $^4$He flow cryostat.
The ab-plane reflectance was measured between 50 and 40 000 \cm\
using an {\it in situ} gold evaporation method~\cite{homes93}. We
estimate the absolute accuracy of the reflectance to be better
than 0.5 \%. For Kramers-Kronig (KK) analysis we used the dc
resistivity \cite{khaykovich04} to extend the data at low
frequency. At high frequencies we used a data of Caimi {\it et
al.}~\cite{caimi04} between 40 000 and 100 000 \cm\ and
extrapolations assuming a free carrier response beyond 100 000
\cm.

\begin{figure}
 \vspace*{-0.4cm}%
  \centerline{\includegraphics[width=3.5in]{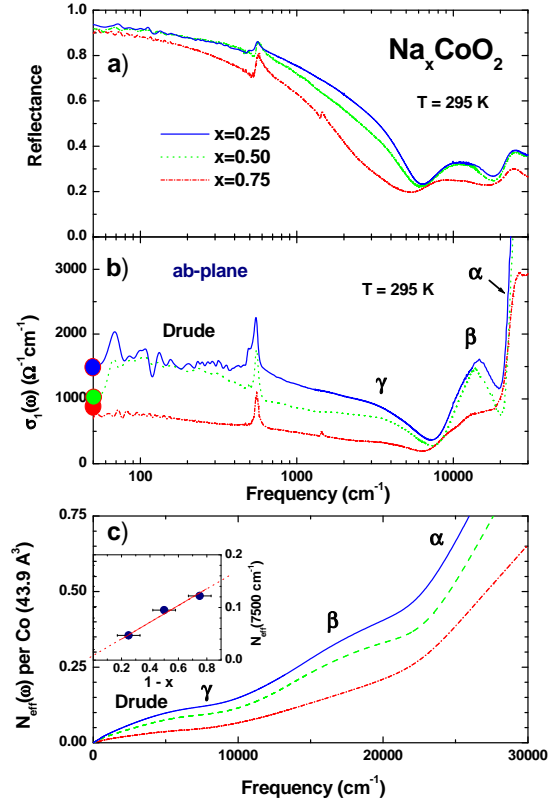}}%
  \vspace*{-0.5cm}%
\caption{\label{fig1} a) The ab-plane reflectance of \Co at three
doping levels, at room temperature. b) The ab-plane optical
conductivity of \Co at the three doping levels at room
temperature. We see a broad Drude-like band at low frequency in
all samples characteristic of an overall metallic conductivity. At
higher frequency there are bands denoted $\gamma$ and $\beta$.
There is a rapid growth of spectral weight of the low lying bands
$\gamma$ and $\beta$ with doping. c) Partial sum rule for the
three doping levels. The Drude weight grows monotonically with
hole doping, shown by the flat region between 5000 \cm\  and 8000
\cm. The inset shows the spectral weight integrated up to 7500
\cm\ with error bars $\pm$ 3 \% as a function of hole doping with
error bars $\pm$ 0.08.}
\end{figure}

Fig. 1a shows the overall reflectance at each of the three doping
levels at room temperature. The reflectance has a characteristic
linear variation with frequency with the slope of the curve
decreasing as the materials become more metallic at the higher
hole doping levels. This is similar to what happens in the
cuprates where the reflectance slope can be used as an internal
standard of the doping level \cite{hwang04}. Using this criterion,
we can see that three different samples show reflectances that
follow the Na doping levels monotonically.

Fig. 1b shows the frequency dependent conductivity at room
temperature at the three doping levels obtained by KK analysis of
the reflectance. The dc resistivity is marked on the ordinate
axis. We see a low frequency Drude-like band but with a high
frequency tail that could be interpreted as an additional
absorption band in the mid-infrared. Following Wang {\it et al.}
\cite{wang03} we will call this the $\gamma$ band. A second
prominent band, the $\beta$ band can be seen at 15 000 \cm\  as
well as a third one $\alpha$ at 26 000 \cm. We see that as the
hole doping level increases there is a strong growth of both the
$\gamma$ and the $\beta$ band intensities as well as the Drude
weight.

The monotonic variation of Drude weight with doping is illustrated
in Fig. 1c where we show the partial spectral weight, obtained by
integrating the optical conductivity, $N_{eff}(\omega_{c})=2 m
V_{Co}/(\pi e^2)\int_{0}^{\omega_{c}}\sigma_{1}(\omega)d\omega$,
where $m$ is the mass of an electron, $V_{Co}$ is the volume per
Co atom, $e$ is the charge of an electron, and $\sigma_{1}$ is the
optical conductivity. The flat region between 5000 to 8000 \cm\
separates the Drude weight from the $\beta$ band weight. The
figure shows that the Drude weight grows monotonically with hole
doping and that the Drude absorption becomes sharper and hence
better defined as the hole doping level increases. Viewed in this
high frequency region and at room temperature, the system becomes
more metallic with hole doping. However, as we will see, at low
frequency and low temperature the two more highly hole-doped
samples ($x=0.50$ and 0.25) are actually insulators. It is
interesting to note that there is no transfer of spectral weight
from the $\alpha,\beta$ and $\gamma$ bands to the Drude component
as the hole doping level is increased as seen in the
La$_{2-x}$Sr$_x$CuO$_{4}$ cuprates \cite{uchida91}.

\begin{figure}
\vspace*{-0.4cm}%
  \centerline{\includegraphics[width=3.5in]{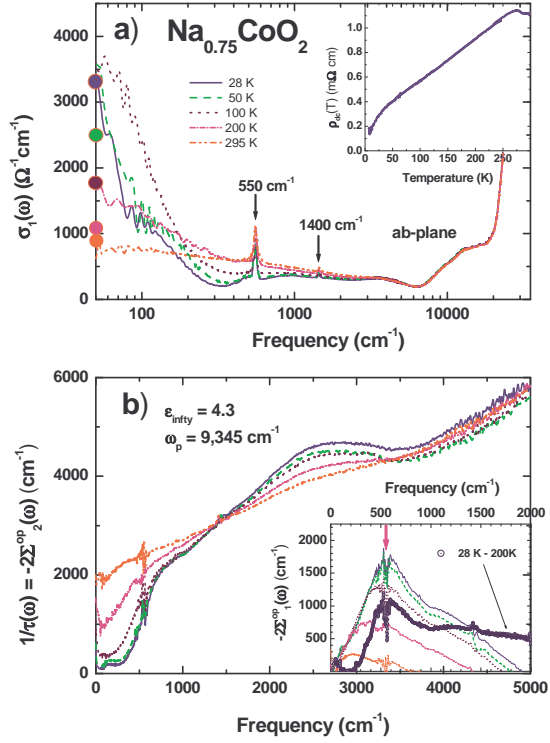}}%
  \vspace*{-1.0cm}%
\caption{\label{fig2} a) The temperature dependence of the
ab-plane optical conductivity of the $x=0.75$ sample and dc
resistivity in the inset. b) The scattering rate $1/\tau$. Here we
subtracted the phonons at 550 and 1400 \cm\ to see the bosonic
mode better \cite{hwang04a}. The inset shows the imaginary part of
the optical scattering rate or the real part of the optical
self-energy. The contribution of the bosonic mode can be separated
from the broad background scattering.}
\end{figure}

The optical properties of the 0.75 sample resemble those of the
underdoped cuprates superconductors. Fig. 2a shows the optical
conductivity of the $x=0.75$ sample at a series of temperatures,
from 295 K to 28 K. At room temperature we see a broad Drude-like
band which splits at low temperature into two components, a
zero-frequency-centered coherent band and a broad band at 1000
\cm\  separated by a slight gap-like depression in the 300 \cm\
region. We also observe two phonons at 550 and 1400 \cm. Drawing
on our experience in the cuprates, we can better illustrate the
physics involved here by plotting the real and imaginary parts of
the scattering rate (or the optical self-energy) using the
extended Drude model\cite{puchkov96,hwang04a}, shown in Fig. 2b
and the inset. Here
$\sigma_{1}(\omega)=\omega_{p}^{2}/[4\pi(\omega-2\Sigma^{op}(\omega))]$,
where $\omega_{p}$ is the plasma frequency and
$\Sigma^{op}(\omega)=\Sigma^{op}_{1}(\omega)+i\Sigma^{op}_{2}(\omega)$
is the complex optical self-energy. Two phonons at 550 and 1400
\cm\  have been removed to allow us to see the bosonic mode
better. Note that the scattering rate is always greater than the
frequency and there are no coherent quasiparticles except perhaps
at the lowest temperatures. We also see that the low-frequency
scattering is dominated by a bosonic mode with an onset frequency
of scattering of the order of 600 \cm.  The contribution of the
mode weakens as the temperature increases and the scattering is
dominated by a strong linear-in-frequency process, very much like
that in the cuprates\cite{hwang04a}. In this picture of a linear
background and a mode, the dc resistivity shows a positive
curvature in the temperature region where the mode becomes
activated \cite{ando04}. The mode and the background can be
separated by plotting the imaginary part of the scattering rate or
the real part of the optical self-energy\cite{hwang04a}, shown in
the inset, along with a difference spectrum where we have
subtracted the 200 K spectrum from the 28 K spectrum. The mode can
be seen as a sharp peak around 600 \cm\  superimposed on the broad
background. This line shape is very similar to what is seen in the
underdoped cuprates \cite{hwang04a}. At low temperatures the the
scattering rate spectrum develops a broad peak at 3000 \cm. This
is reminiscent of the mid-infrared absorption seen in the single
layer cuprates at low doping levels \cite{uchida91}.

\begin{figure}
\vspace*{-0.4cm}%
  \centerline{\includegraphics[width=3.5in]{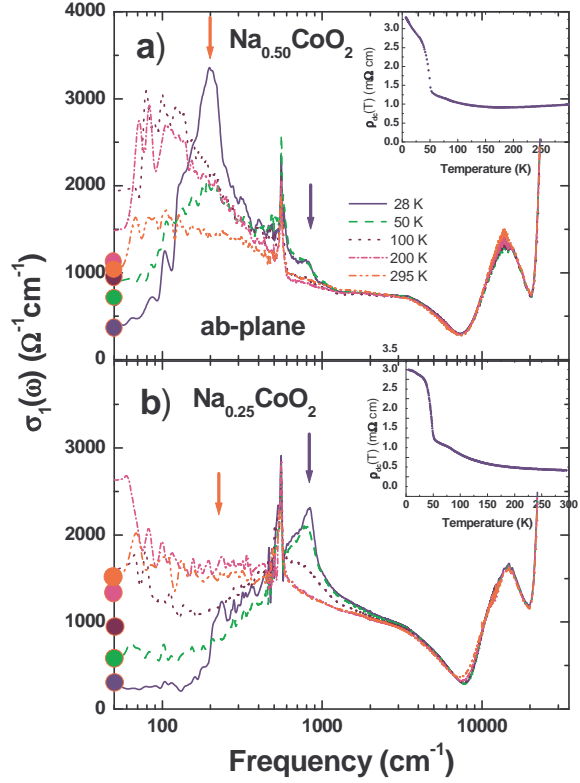}}%
  \vspace*{-1.0cm}%
\caption{\label{fig3} a) Temperature dependence of the optical
ab-plane conductivity of the $x= 0.50$ sample and its dc
resistivity in the inset. b) The optical conductivity of the
$x=0.25$ sample and its dc resistivity in the inset. The symbols
on the vertical axis denote the dc conductivity of the samples. We
see that when the temperature is lowered both samples become more
insulating due to the formation of a gap in the spectrum. The
spectral weight removed from the gap is transferred to the peaks
at higher frequencies delineated by arrows.}
\end{figure}

Fig. 3a shows the frequency dependent conductivity of the $x=0.50$
sample at various temperatures. We show the dc conductivity on the
zero frequency axis as well. The dc resistivity shows insulating
behavior with the 53 K and 88 K anomalies (see inset of Fig. 3a).
We notice that there is an overall agreement between the lowest
measured infrared conductivity and the dc conductivity. This is
evidence of two important facts. First, the samples are reasonably
homogeneous without any channels and inhomogeneities that might
short circuit the current. Second, there are no low-lying
collective modes in the spectral region between zero frequency and
50 \cm, our lowest measured frequency. This is not always the case
in strongly correlated systems. In the one-dimensional organic
conductors there is a large discrepancy between the highly
metallic dc conductivity and nearly insulating gap-like behavior
in the infrared, a signature of sliding charge-density wave(CDW)
transport\cite{ng84}. There is no evidence for such effects in our
samples of \Co, although is should be noted that there is about a
factor of two discrepancy between the dc values and the peak that
develops around 100 \cm. This is similar to what is seen in the
cuprates at low frequency and has been attributed to localization
effects \cite{basov98}. There is evidence of similar localization
effects in the $x=0.75$ sample where there is also a discrepancy
at low temperature between the dc and the lowest far infrared
measurement and the dc conductivity. In all cases the dc value is
{\it lower} suggesting localization, whereas in the CDW systems
with sliding waves, the dc value is {\it higher} by a substantial
factor.

This system shows very interesting temperature dependent properties
at low frequencies as expected from the temperature dependence of the
dc resistivity. There is a sharpening of the free carrier peak as the
temperature is lowered from 295 to 200 K. Between 200 and 100 K the
spectra do not change much. Below 100 K a gap develops rapidly and is
accompanied by a sharp peak at 200 \cm. At our lowest temperature 28 K
the gap seems fully developed with a width 110 \cm. In agreement with
this overall energy scale, the spectral weight lost in the gap region
is recovered by 600 \cm. These changes in the optical properties are
clear evidence of a low temperature pinned density wave phase with a
resonance frequency 200 \cm.

Fig. 3b shows the conductivity of the most highly hole doped
sample with $x=0.25$ at different temperatures. Again, there is
reasonable agreement between the dc and infrared magnitudes of the
conductivity suggesting insulating behavior and the absence of
sliding density waves. The dc resistivity also shows insulating
behavior with 53 K and 88 K anomalies (see inset of Fig. 3b).
Similar 52 K anomaly has been observed in a highly hole doped and
fully hydrated sodium cobaltate,
Na$_{0.30}$CoO$_{2}$$\cdot$1.4H$_2$O~\cite{jin03}. The origin of
the anomaly in the dc resistivity of high hole doped systems is
not clear yet.

The gap-like depression of conductivity below 200 \cm\ in the
$x=0.25$ sample starts to develop between 200 and 100 K and is
nearly fully formed at 28 K, our lowest measured temperature. Like
the pseudogap seen in c-axis transport in the cuprates, the gap
here does not close at temperature but has a constant width and
fills in as the temperature is raised. There is a notable sharp
peak at 800 \cm\ that grows in parallel with the gap. It starts
below 200 K and has its most rapid rate of increase between 100 K
and 50 K. There is a transfer of spectral weight from the gap to
the peak, and by 1000 \cm, nearly all the spectral weight lost in
the gap region has been recovered. The size and the onset
temperature of the gap, the center frequency of the sharp peak,
and the spectral weight recovering frequency in the $x=$0.25
sample are higher than the corresponding ones in the 0.50 sample.
These factors strongly suggest that there is a distinct and
separate pinned CDW insulating phase in the 0.25 system.

It is interesting to note that in the 0.50 sample there is a weak
remnant of the 800 \cm\  mode suggesting that the sample may have
an admixture of the 0.25 phase, about 15 \% based on the spectral
weight of the peaks. Similarly, it is possible that the feature
230 \cm\ seen in the 0.25 sample may be related to a minority
phase of the 0.50 peak at 200 \cm, about less than 2 \% based on
the spectral weight of the peaks. The peaks are denoted by arrows
in the figures.

The only transverse optic phonon that can be seen in the spectra
is a sharp peak at 550 \cm. Its spectral weight ($\omega_{p} =
870$ \cm) is consistent with what is expected for a phonon that
involves oxygen motion. In the $x=0.75$ crystal there is an
additional phonon mode at 1400 \cm\ that we attribute to a small
amount of carbonate, one of the starting materials of the
synthesis process.  The 550 \cm\ mode in the $x=0.50$ sample
appears to split between 100 K and 50 K. The 550 \cm\ mode in 0.75
sample does not show any splitting as temperature is lowered. The
550 \cm\ mode acquires a slight Fano shape below 100 K at the
highest hole doping levels.  We do not observe, in any of our
samples, the dramatically enhanced spectral weight of phase
phonons associated with CDW order seen in the organic charge
transfer salts \cite{rice76}.

There is an alternate explanation of our results on the two insulating
samples in terms of two different pinning sites for the charge density
waves with the stronger pinning potential dominating at the higher
doping level. However, to explain the fact that we have two insulating
samples with Drude weight that differs by 20 \% we have to assume that
the insulating region in the phase diagram at $x=0.5$ is quite wide.
Another possibility that could account for the extra spectral weight
in the 0.25 sample is a two-phase system with metallic drops present
in the insulating host.  This can be ruled out since such drops would
not add to the Drude weight but give a high frequency band at
$\omega_p/\sqrt{3}$.

In summary the optical conductivity of \Co has many common
elements with the cuprate high temperature superconductors but
there are also striking differences. The free carrier absorption
is dominated by a continuum of scattering processes that extend to
nearly 0.75 eV in energy. For the 0.75 sample, as the temperature
is lowered, a bosonic mode appears, very much like the mode that
has been attributed to the magnetic resonance in the cuprates
\cite{carbotte99,hwang04a}. The mode even has a frequency close to
that of the cuprates.

The two samples with the  higher hole doping levels ($x=$ 0.50 and
0.25) become insulators at low frequency and low temperature as
results of the development of two different size gaps and pinning
resonance peaks that are quite a bit higher than similar features
seen in the cuprates\cite{timusk95,dumm02,benfatto03}. There are
localization effects in the cuprates, but these are hard to
observe because of the appearance of the superconducting
state\cite{boebinger96}. In contrast, in \Co the charge ordering
potentials are much stronger giving rise to clear gaps in 0.50 and
0.25 materials as well as absorption peaks at quite high
frequencies. These may be processes that compete with
superconductivity and may be the cause of the absence of
superconductivity of \Co or the relatively low superconducting
transition temperatures of \Co system with additional H$_{2}$O
layers. On the positive side, it seems that the basic attractive
force in the two systems is very similar suggesting that if the
competing charge order can be suppressed by the manipulation of
the Na lattice a higher superconducting $T_c$ may be possible in
this system.

\acknowledgments We thank Patrick Lee and Takashi Imai for
encouragement and helpful discussions. The crystal growth at MIT
was primarily supported by the MSEC program of the National
Science Foundation under award number DMR-02-13282.

\end{document}